**Imaging Nonlinear Spin Waves in Magnetoacoustic Devices**

N. Beaver[1], B. Luo[2], S-W. Chiu[3], D. A. Bas[4], P. J. Shah[4], A. Franson[4], M. S. Wolf[4], M. R. Page[4], M. J. Newburger[4], L. Caretta[3], N. X. Sun[2], P. Stevenson[1]

[1]Department of Physics, Northeastern University, Boston MA
[2]Department of Electrical Engineering, Northeastern University, Boston MA
[3]School of Engineering, Brown University, Providence RI
[4]Materials and Manufacturing Directorate, Air Force Research Laboratory, Wright-Patterson Air Force Base, Ohio

**Abstract:** Magnetoacoustic systems offer promising platforms for next-generation sensors and computing applications, but understanding their nonlinear dynamics remains challenging. Here, we use nitrogen vacancy (NV) centers in diamond to spatially map nonlinear magnon scattering processes in FeGaB/LiNbO$_3$ magnetoacoustic devices with sub-micron resolution. We observe highly heterogeneous magnetic noise generation under acoustic driving at 1425 MHz, with responses varying dramatically across micron length scales. Time-domain measurements reveal threshold-like nonlinear behavior where NV center spin relaxation rates increase over two orders of magnitude as drive power is increased. These findings reveal microscopic noise sources that limit magnetoacoustic sensor performance while simultaneously demonstrating how acoustic mode engineering could enable selective control of nonlinear magnon processes.

**Introduction:** Hybrid material platforms — systems with multiple, distinct degrees of freedom coupled together — offer many technological advantages, such as increased tunability, novel functionalities, and improved efficiency[1–3]. Among this class of materials, magnetoacoustic systems (where acoustic and magnetic degrees of freedom are coupled) have shown outstanding promise as sensors[4,5], distributed communication elements[6,7], and novel computational platform[8]. Moreover, they are able to leverage the long decay length found in *e.g.* LiNbO$_3$ to generate coherent magnetic excitations on length scales that are orders of magnitude greater than those found in typical metallic ferromagnets[9]. However, with additional degrees of freedom comes additional complexity; understanding and mitigating noise processes in these systems presents a significant experimental challenge.

In parallel, there is a growing need for compact, power-efficient systems which can generate strong nonlinearities to enable next generation computing applications, such as neuromorphic computing and other non-von Neumann approaches. These applications require systems where a strongly non-linear response can be realized; magnetic excitations (spin waves) have already shown great promise in this area[10,11], but there remain many outstanding materials challenges such as sample heterogeneity and fast spin wave damping.

These two applications motivate the need to better understand nonlinear magnetic processes in magnetoacoustic systems. Nonlinear interactions can act as a source of noise, degrading sensor performance[12], where understanding them more fully — and where they differ from conventional excitation methods — would enable physically-inspired design principles to be developed. On the other hand, if these nonlinearities are to serve as a component of next-generation computing then engineering their emergence at low powers is of vital importance; again, understanding the physical mechanism represents an important step towards optimized devices.

**Imaging Spatial Heterogeneity In FeGaB:** To probe the dynamics of acoustically-driven magnetic excitations, we use the nitrogen vacancy (NV) center in diamond. This fluorescent defect is able to characterize magnetic fields from DC-GHz frequencies with high spatial resolution and sensitivity[13,14]. As a model magnetoacoustic system, we use FeGaB bilayer films (FeGaB[20nm]|$Al_2O_3$[5nm]|FeGaB[20nm]) on $LiNbO_3$, previously reported to have low coercive fields, is self-biased, and has strong magnetoelastic coupling[15] and non-reciprocal behavior[16]: all essential elements of high-performance sensors. Moreover, the combination of large magnetoelastic response and low damping makes this an excellent platform to explore nonlinear dynamics.

To probe the magnetic dynamics in this system, we use a single crystal diamond with a dense, uniform ensemble of NV centers 30nm below the surface formed by ion implantation (Quantum Diamonds) as our sensor. This is placed on a FeGaB bilayer film patterned on top of a $LiNbO_3$ surface acoustic wave (SAW) device which is excited at its 5$^{th}$ harmonic (1425 MHz). This experimental configuration was chosen because of previous demonstrations of exceptionally strong magnetoelastic responses in this system[16,17]. We observe a resonance at 1425 MHz in both time-gated electrical transmission measurements and in our NV center optically-detected magnetic resonance (ODMR) experiments. However, our ODMR-detected linewidth is significantly smaller than our electrical measurements (4MHz, consistent with previous measurements[18], vs 1.5 MHz for ODMR-detection, Figure 1). As we discuss later, we attribute this to the highly nonlinear processes being detected by our NV center; we are primarily sensitive to spin wave processes which only occur above a threshold drive power, resulting in an apparent decrease in our linewidth.

We observe a remarkably heterogeneous response across our sample with a strong dependence on the external magnetic field applied (Figure 2). In different regions, we observe variations in both the magnitude and spatial profile of the response; in Region 1, we observe distinct spatial features on the order of $10 \mu m$ at external bias fields of 11.5 G, while in Region 2 we observe smaller features which also display a clear magnetic field dependence (Figure 2 and extended datasets in SI). These features disappear as the magnetic field is increased and are persistent over multiple cycles of increasing and decreasing the magnetic field. We verify with longitudinal magneto-optical Kerr effect (MOKE) imaging that this is not due to the presence of static magnetic domains in our films (see SI). In our regions of interest, we observe uniform magnetization, consistent with the extremely low coercive fields of FeGaB (<5G) measured; this suggests a more subtle origin to these features, arising from the dynamic effects which we discuss later.

Plotting the average response of the field of view (Figure 2c) for each frequency yields a curve which agrees well with the conventional acoustically-detected magnetic resonance (ADMR) measurements on this sample for Region 1 but yields a much broader response for Region 2. In both cases there is a strong NV center response at the same magnetic field as the strongest ADMR response. However, if we analyze the magnetic field dependence locally, we find Region 2 shows significant spatial heterogeneity; if we instead analyze select regions we see much narrower responses with magnetic field, highlighting the effect of heterogeneity. We also analyze the angular response of Region 1, finding that both the NV-detected response and the conventional ADMR measurements show responses peaked around the same angle, further confirming that our NV-detected signal is related to acoustic driving of magnons. However, we observe a broader angular dependence in the NV-detected response. This, combined with the heterogeneous response we see, motivates further exploration of the mechanism of coupling between the magnetoacoustic device and the NV center.

**Spin Relaxation Driven by Magnetoelastic Excitations:** The magnetic field dependence of the conventional ADMR signal is well understood; changing the magnetic field brings spin wave modes which share the same wavevector $k$ as the acoustic mode into resonance with the SAW transducer frequency, resulting in efficient transduction between acoustic excitations and spin waves. We confirm that the response we observe with the NV centers arises from these magnetoelastic interactions in two ways: first, through the agreement between the spatially averaged NV data and the ADMR data, and second the lack of response over the bare $LiNbO_3$ substrate (see SI).

However, the response we observe with the NV center deviates from the ADMR in several important ways: the linewidth is substantially narrower, there is significant spatial heterogeneity, and the angular dependence is broader. To explore the mechanism of action in detail we turn to time-domain experiments. First, a green laser pulse polarizes our NV center ground state spin, then a separate microwave pulse excites the magnetoacoustic response, finally, a final green laser pulse probes the effect of this on the NV center spin, summarized in Figure 3a. Here, we observe an exponential decay of spin polarization as the duration of the microwave drive increases (Figure 3b), indicating incoherent relaxation of the NV center spin driven by magnetic noise induced by the microwave; we see no evidence of coherent oscillations indicative of coherent driving. Moreover, we find the drive-induced relaxation rate has an extremely nonlinear power dependence, displaying threshold-like behavior where the rate increases by >100x as the power is tripled.

The increase in $T_1$ relaxation rate is responsible for the ODMR contrast observed in Figure 1. Changes in the NV center $T_1$ change the steady state populations of the different spin projections under continuous excitation, which in turn changes the steady state photoluminescence[19,20], consistent with the similar power dependences observed for the continuous wave (CW) ODMR data and $T_1$ relaxation rate. Connecting the $T_1$ time and ODMR contrast thus reveals that the magnetic noise under microwave excitation is spatially inhomogeneous in our device.

From our pulsed measurements, we are also able to exclude other contributions to the CW ODMR signal; the ground state splitting of the NV center is 2.87GHz in the absence of external magnetic fields, meaning we do not expect to directly drive the ground state transition. The excited state of the NV center has a zero-field splitting of ~1420MHz and a linewidth of ~100MHz[21]; however, we exclude the possibility of coherently driving the excited state spin because we observe the same frequency response in pulsed measurements, where there is no population in the excited state when the microwave drive is applied.

**Origin of the Magnetic Noise Response:** Our data reveal that the acoustic drive at 1425MHz creates an increase in the spectral density of magnetic noise in the >2.7 GHz range (the ground transition energies of the NV center) which varies nonlinearly with power. While there are many possible processes which can generate broadband magnetic noise – several of which have been previously characterized using NV centers[22–25] – the details of our results are inconsistent with many of these and require a new mechanism. We briefly review these processes and how our data enables us to exclude these as explanations:

*Temperature changes:* Under our microwave excitation conditions, device heating can occur at high microwave powers (see SI for characterization). In principle, a temperature change will increase the population of thermally excited spin waves[23]. However, in the high temperature limit this contribution to the spectral density is governed by the Rayleigh-Jeans distribution, $S(\omega) \propto kT/\hbar\omega$. To explain the hundred-fold change in relaxation rate from this effect, we would require a temperature of >30,000K. Thus, we ignore the effect of drive-induced temperature changes on the density of states in our analysis.

*Changes in Spin Chemical Potential:* Drive-induced spin decay with threshold-like behavior has been observed previously in Yttrium Iron Garnet (YIG) films excited at spin-wave mode resonances, where changes in the spin chemical potential result in an increased noise spectral density at the NV center resonance[23]. Here, the spin chemical potential can be defined as $\mu = \hbar\omega_{NV}\left(1 - \frac{\Gamma_0}{\Gamma_{MW}}\right)$, where $\Gamma_i$ is the relaxation rate in the presence ($\Gamma_{MW}$) or absence ($\Gamma_0$) of the microwave drive and $\omega_{NV}$ is the NV center resonance frequency. The maximum allowed value for this chemical potential is equal to the band minimum[26]. From our data (Figure 3b) we would extract $\mu/\hbar > 2.8$GHz, which is significantly larger than the ferromagnetic resonance frequencies under our experimental conditions[16,17] and thus not a physically meaningful solution.

*Second harmonic generation:* Harmonics of the drive frequency are commonly observed in magnetic materials under strong driving[27–31]. However, this is explicitly a non-threshold behavior[31], where the observed amplitude varies quadratically with input drive (and also SI). Moreover, simple second harmonic generation fails to explain much of our observed response. Specifically, the significant spatial inhomogeneity and the large (>15%) magnitude contrast response observed. In an ensemble of NV centers, the observable PL contrast for each orientation is approximately one-quarter of the contrast for an individual NV center, giving rise to typical ensemble contrasts on the order of 5%; thus, multiple orientations of NV centers are being addressed in our experiment, which requires a broad frequency response.

*Four-magnon scattering:* Brillouin light scattering measurements have shown substantial spectral broadening with threshold-like power dependence arising from four-magnon scattering processes[32], where driven spin wave modes interact to yield new excitations constrained by $\omega_1 + \omega_2 = 2\omega_{RF}$. This effect has also been observed previously in magnetoacoustic systems[33]. However, the maximum broadening possible is limited by the difference in energy between the drive and the minimum spin wave frequency. In our experiments, we drive at approximately the band minimum; thus, broadening from four-magnon scattering cannot explain the wideband response we see at >2.8GHz.

*Multi-step processes:* To explain our data, we instead consider two alternative scenarios. First, we consider a modification of the four-magnon scattering process, where we also consider the effects of thermally excited magnons. Here, a higher energy thermally-excited mode may combine with our drive frequency to yield two magnons at the NV center's frequency. This effect has previously been observed in the ferromagnetic $Ni_{0.65}Zn_{0.35}Al_{0.8}Fe_{1.2}O_4$ films[34] and could explain the spectral characteristics we see: broadband responses at >2.8GHz with a threshold power dependence. However, this mechanism does not provide any explanation for the spatial heterogeneity we observe.

As an alternative explanation, we consider two three-magnon processes acting in concert: magnon confluence (where two spin waves with opposite wavevectors combine to yield a low $k$ mode at double the frequency) and a first-order Suhl instability, where a low $k$ mode is driven above a threshold. In this picture, which we justify in the following section, our acoustic drive generates spin wave excitations with a well-defined wavevector $k$, which interacts with opposite $k$ modes generated by either acoustic scattering or thermal excitation to generate spin waves at $\omega_{conf} = 2\pi \times 2850 MHz$. By itself, this process cannot explain the large response we observe (see *Second harmonic generation* discussion). However, the population of this mode increases with power until above a steady-state threshold it undergoes a Suhl-instability process generating modes of opposite wavevector. Importantly, these modes need not be the same as used to initially drive the confluence process, but they can seed future rounds of the magnon confluence process. This cycle is shown schematically in Figure 4. Each roundtrip through this process can change the frequency and wavevectors involved in a non-phase-conserving manner, generating broadband, incoherent noise at the NV center frequency band.

First, we demonstrate that this process is allowed by the dispersion of our FeGaB bilayer structure. Given the dimensions of our film, we neglect the thickness modes which occur at much higher frequencies. We use the previously reported parameters to perform micromagnetics simulations using MuMax3 to calculate the dispersion of the material, finding good agreement between these numerical calculations and previous analytical models[17,35]. Because of the bilayer nature of our sample, at each $k$ we expect two modes, though the upper branch quickly moves beyond the range of our experiment even for small $k$, as shown in Figure 4a. This provides the basis for our proposed confluence process; we drive $k$ spin waves on the lower branch with our acoustic excitation, which can combine to yield a low $k$ mode on the upper branch (Figure 4b).

To assess the feasibility of this potential mechanism, we perform micromagnetic simulations, where we break the simulations into two parts to simplify the analysis, and use drive parameters estimated from experimental measurements (see SI). First, we find that under a drive with non-zero $k$, we can indeed generate responses at $2\omega$ and recover the expected $B_{drive}^2$ dependence. Next, we drive the system at $2\omega$ with a uniform ($k = 0$) excitation to probe the instability process. Here, we expect to underestimate the extent of this process; in principle, the uniform mode can decay to any pair of magnons which satisfy $k_1 + k_2 = 0$ and $\omega_{uniform} = \omega_1 + \omega_2$ with nontrivial phase relationships, potentially leading to destructive interference. Nevertheless, we are able to observe threshold-like behavior when driving the uniform mode, albeit with less pronounced nonlinearity than observed in our experiments.

Finally, we simulate the response of our system to driving with two plane-wave excitations: one fixed at the experimental direction of our propagating acoustic wave in our experiment, and a second whose angle we vary. As shown in Figure 4e, this configuration can generate responses at $2\omega$, $k \approx 0$, but is extremely sensitive to the propagation direction of the second source, peaking at $k_2 \approx -k_1$, as expected for a magnon confluence process.

Since our SAW device generates acoustic waves with a well-defined $k$, the question naturally arises of the origin of this second wavevector. Reflections from the receiver electrodes are one possible source, as are other sources of acoustic scattering from the film. This would provide an explanation for the substantial heterogeneity we see in the noise response (Figure 2); as noted earlier, MOKE imaging reveals that the magnetic film is uniformly magnetized in the regions we probe, ruling out one obvious source of heterogeneity in our experiments. However, these same imaging techniques reveal the presence of topographical growth defects — a potential acoustic scattering site — near the region we observe, suggesting acoustic scattering may be responsible for the dramatic variations in drive-induced noise we observe.

Other multistep nonlinear mechanisms could also explain some of our data; for example, four-magnon scattering could generate a spectrally-broad response centered at 1425 MHz, which could then scatter to generate a response at double the frequency. This would be consistent with the spectral response we expect, and would also have a threshold power-dependence; however it would not provide an explanation for the spatially-inhomogeneous response we observe. The mechanism we propose to explain our data is intrinsically sensitive to the wavevector of the spin wave modes involved, highlighting the importance of considering scattering processes not only in the magnetic layer, but also in the piezoelectric medium.

## Conclusions

Using a diamond-based quantum sensor, we are able to image nonlinear magnon scattering processes in a hybrid magnetoacoustic device, revealing a highly heterogeneous spatial response. Through a combination of power-dependent, time-domain measurements and micromagnetic simulations, we propose a two-step mechanism involving magnon confluence and instability which explains the threshold-like power dependence we observe and suggests acoustic scattering plays a role in the spatial response of the system. These results provide new insight into the microscopic mechanisms which limit existing magnetoacoustic sensors, suggesting acoustic scattering should be viewed as a potential noise source in these devices. However, the rich range of nonlinear behaviors we observe also suggests that shaping the acoustic modes – using, for example, topologically protected modes – is a route to selectively engineering specific nonlinear processes through wavevector selection.

## Acknowledgements

P.S. and N. X. S acknowledge support from Northeastern University's TIER 1 Internal Seed Grant Program. MOKE imaging at Brown was supported under AFOSR FA9550-24-1-0169. This work is partially supported by the Air Force Office of Scientific Research (AFOSR) Award No. FA955023RXCOR001

**Figures:**

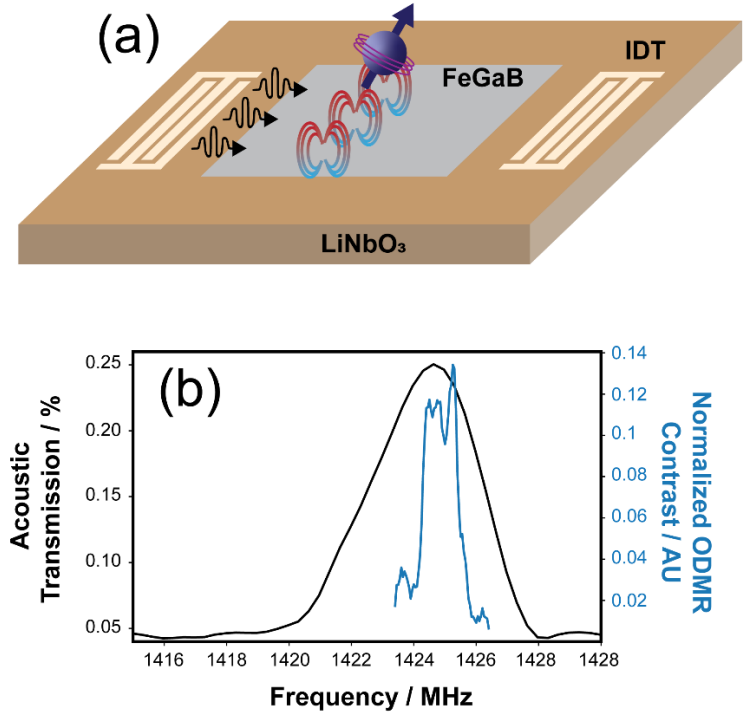

**Figure 1 –** (a) Schematic of the central concept: interdigitated transducers (IDTs) generate acoustic waves (black), which in turn generate spin waves (red/blue dipoles) in a magnetoelastic material (FeGaB). A spin sensor — the nitrogen vacancy center — detects the magnetic fields generated. (b) Comparison of time-gated transmission of the acoustic wave measured by a vector network analyzer (black) and the optically-detected magnetic resonance of the NV center (blue).

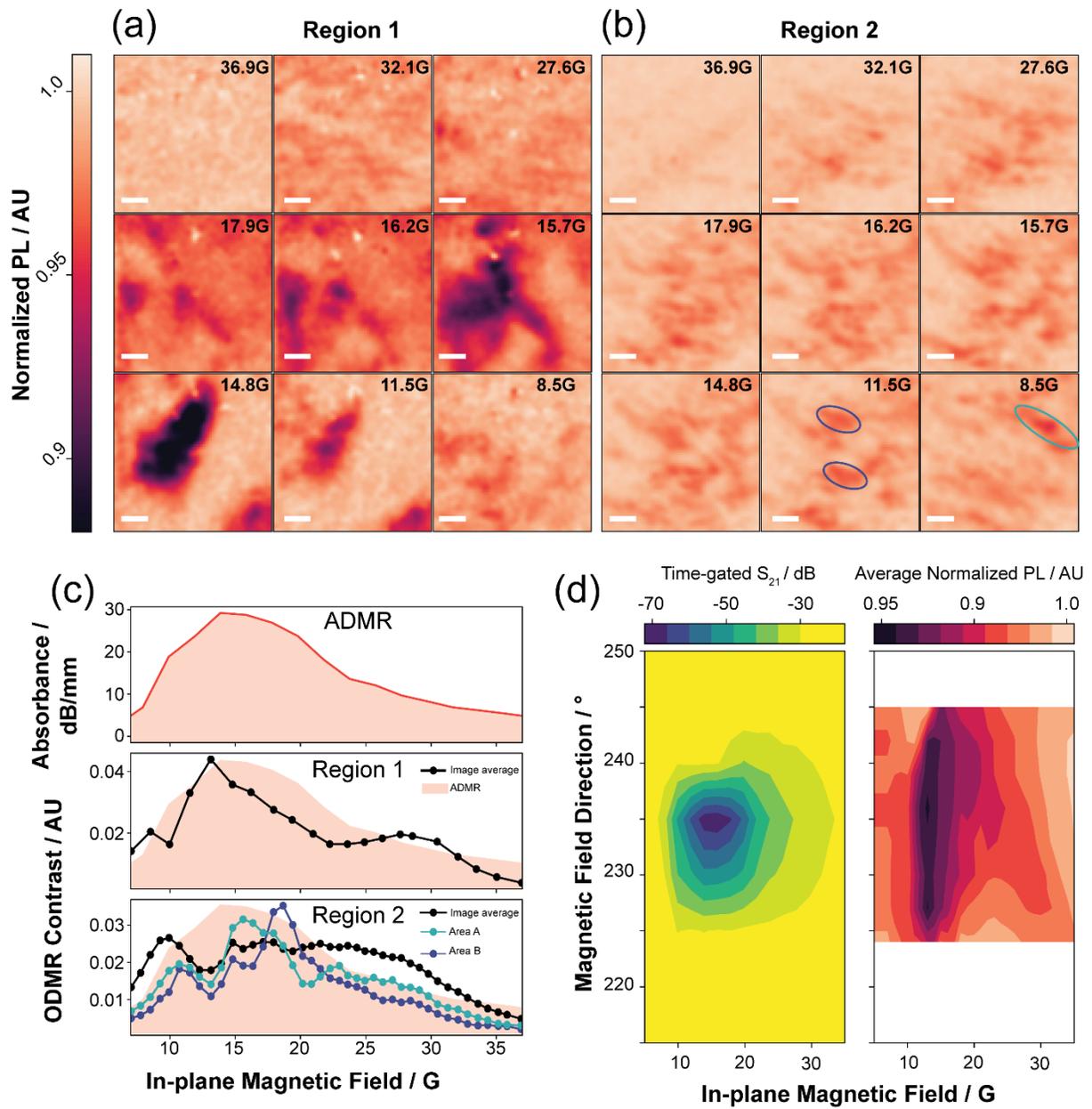

**Figure 2 –** (a) Spatial maps of the NV center response show substantial heterogeneity. (b) A second region shows a similar field-dependent response, but with a different spatial pattern of the response. Scale bar is $5\mu m$ (c) Comparison of the field-dependent response measured with the NV center and detected with conventional ADMR responses. (d) The angular response of the time-gated electrical transmission measurements ($S_{21}$) show a response consistent with the angular dependence of the NV center ODMR.

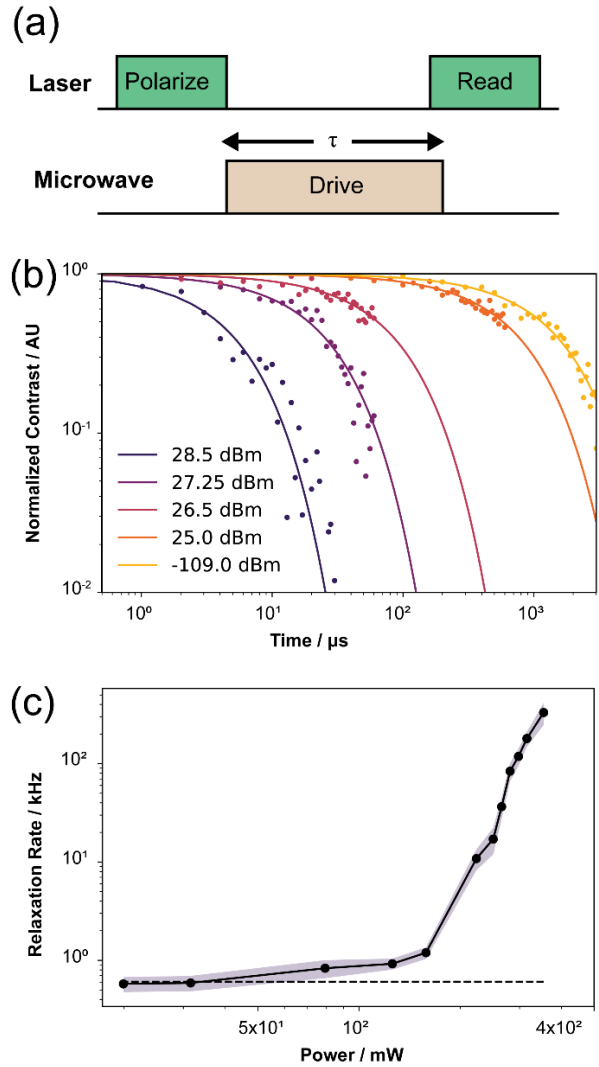

**Figure 3 –** (a) Pulse sequence used to measure the increased spin-lattice relaxation rate caused by microwave driving. (b) The microwave drive causes fast (incoherent) relaxation of the NV center spin state. (c) The power dependence of the relaxation rate is highly nonlinear. The shaded area shows the 95% confidence interval of the points. The dashed black line shows the relaxation rate with no microwave drive.

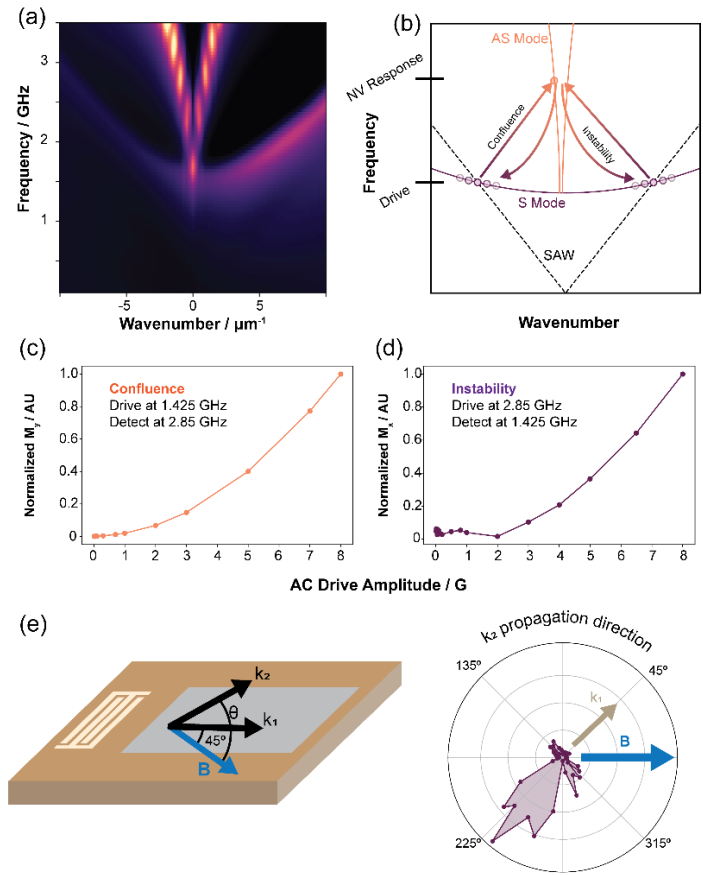

**Figure 4 –** (a) Simulated dispersion of the FeGaB bilayer for wavevectors aligned with the SAW propagation direction. (b) Concept for the proposed mechanism: two spin waves combine to generate a higher frequency mode (confluence); above a threshold, this splits into two similar-frequency modes, and the cycle continues. (c) Micromagnetic simulations of the confluence process, showing the expected $B^2$ dependence. (d) Micromagnetic simulations of the instability process, which yields a negative intercept when fit with quadratic or linear power dependences, indicating a threshold-process. (e) Magnitude of the $k = 0, \omega = 2\pi \times 1.425$GHz response as a function of the angle of the second spin wave, showing the highly anisotropic response. The directions of the first mode and the static magnetic field are indicated.